# Artificial Intelligence in Mental Health and Well-Being: Evolution, Current Applications, Future Challenges, and Emerging Evidence

# (A Short Review)

Hari Mohan Pandey

Department of Computing and Informatics, Bournemouth University, United Kingdom

profharimohanpandey@gmail.com

**Abstract:** Artificial Intelligence (AI) is a broad field that is upturning mental health care in many ways, from addressing anxiety, depression, and stress to increasing access, personalization of treatment, and real-time monitoring that enhances patient outcomes. The current paper discusses the evolution, present application, and future challenges in the field of AI for mental health and well-being. From the early chatbot models, such as ELIZA, to modern machine learning systems, the integration of AI in mental health has grown rapidly to augment traditional treatment and open innovative solutions. AI-driven tools provide continuous support, offering personalized interventions and addressing issues such as treatment access and patient stigma. AI also enables early diagnosis through the analysis of complex datasets, including speech patterns and social media behavior, to detect early signs of conditions like depression and Post-Traumatic Stress Disorder (PTSD). Ethical challenges persist, however, most notably around privacy, data security, and algorithmic bias. With AI at the core of mental health care, there is a dire need to develop strong ethical frameworks that ensure patient rights are protected, access is equitable, and transparency is maintained in AI applications. Going forward, the role of AI in mental health will continue to evolve, and continued research and policy development will be needed to meet the diverse needs of patients while mitigating associated risks.

## 1. Introduction

Artificial Intelligence (AI) in mental health care represents a transformative development in the delivery of psychological support and treatment, utilizing advanced technologies to address an array of mental health challenges, including anxiety, depression, and stress disorders. As mental health issues continue to escalate globally, AI applications have emerged as innovative solutions, enhancing accessibility, personalizing treatment, and providing real-time monitoring that can significantly improve patient outcomes.[1] [2] This intersection of technology and mental health care is notable not only for its potential to alleviate existing barriers to access but also for the ethical implications and challenges it raises, including concerns about privacy, data security, and algorithmic bias.

Historically, the use of AI in mental health has evolved from early chatbot models, such as ELIZA in the 1960s, to contemporary machine learning systems that analyse complex datasets for diagnostic and therapeutic purposes [3][4]. The increasing recognition of mental health as a critical component of overall well-being has catalysed the integration of AI technologies [5][6], prompting both healthcare providers and researchers to explore their applications across various settings, from clinical environments to mobile health platforms.[7][3] However, this rapid advancement has also unveiled a spectrum of challenges, including the need for rigorous ethical frameworks and guidelines to safeguard patient rights and ensure equitable access to these emerging resources.[8][9] Controversies surrounding AI in mental health primarily centre on issues of trust, safety, and efficacy. Critics argue that the lack of transparency in AI algorithms, coupled with potential biases in data handling, could exacerbate existing inequalities in mental health care delivery.[8][4] Furthermore, ethical dilemmas related to patient privacy and informed consent remain pressing concerns as mental health applications increasingly rely on sensitive user data for functionality.[10][4]Consequently, ongoing research and policy development are crucial to navigating these challenges and ensuring that the integration of AI into mental health care is conducted responsibly and ethically, prioritizing the needs and rights of patients.[11] As the landscape of mental health care continues to evolve, AI is poised to play a pivotal role in shaping how individuals access support and treatment, ushering in a new era of personalized, data-driven interventions. Emphasizing the importance of equitable practices and robust ethical standards will be essential in harnessing the full potential of AI technologies in mental health, ultimately fostering a more inclusive and effective system of care for all individuals.[12][11]

The remainder of the paper is organized as follows: **Section 2** provides an overview of the historical background of AI in mental health, tracing its evolution and early applications. **Section 3** highlights the current applications of AI in mental health and well-being, focusing on its diverse uses in diagnosis, treatment, and patient support. **Section 4** discusses the benefits of integrating AI into mental healthcare, emphasizing improvements in accessibility, early detection, and personalized interventions. **Section 5** outlines the challenges and limitations associated with AI in mental health, including ethical concerns and technological barriers. **Section 6** presents key evidence and research findings related to the use of AI in mental healthcare, illustrating its effectiveness and potential. Finally, **Section 7** offers concluding remarks, summarizing the key insights and future directions for the field.

## 2. Historical Background

### 2.1 Initial Applications of AI in Mental Health

AI's application in mental health care has evolved over the past few decades, from simple algorithms to sophisticated machine learning models capable of predicting treatment outcomes and classifying mental illnesses based on extensive datasets [1][6]. Early efforts primarily focused on developing chatbots and basic computer programs that could mimic therapeutic conversations, such as the famous ELIZA program in the 1960s, which was designed to simulate human interaction [5][1]. These early models set the groundwork for more advanced applications, paving the way for innovations in therapy delivery and patient monitoring.

### 2.2 Evolution of Mental Health Care

The history of mental health care is marked by significant shifts, particularly in the perception and treatment of mental disorders. For much of history, individuals with mental illnesses were often marginalized, subjected to prejudicial practices based on their identity, which has left enduring impacts on contemporary clinical practice [13]. This historical backdrop underscores the importance of developing new models of care that prioritize equity and do not perpetuate past harms.

### 2.3 Emergence of Artificial Intelligence in Health Care

Artificial intelligence (AI) began to gain traction in various healthcare fields, including oncology and radiology, in the late 20th century. However, its integration into mental health care has been relatively modest [14]. The recognition of the high morbidity and mortality associated with psychiatric disorders has intensified the need for innovative solutions. As a response, AI has started to emerge as a viable tool to address the growing mental health crisis, particularly as traditional care models prove insufficient in meeting increasing demands [2].

### 2.4 Recent Developments and Trends

In recent years, the surge of interest in AI technologies has coincided with a broader recognition of mental health as a critical component of overall well-being, particularly as disorders like depression and anxiety have emerged as leading causes of global disability [3]. Modern AI systems are now being explored for their potential to augment traditional mental health care through predictive analytics, personalized treatment plans, and the automation of therapeutic interventions [7][6]. This represents a significant shift toward integrating technology in mental health services, particularly in response to the ongoing shortage of mental health professionals and the increasing stigma surrounding mental illness [2]. The evolution of AI in mental health care highlights both the challenges and opportunities that lie ahead. As the landscape continues to change, it is imperative to ensure that these technologies are implemented thoughtfully, addressing the historical biases and inequities that have shaped mental health care13].

## 3. Current Applications

### 3.1 Overview of AI Applications in Mental Health

The integration of artificial intelligence (AI) into mental health care has given rise to numerous applications designed to address various mental health challenges, including anxiety, depression, and stress. Notable examples include Wysa, Woebot, Elomia, Youper, Koko, and Replika.[2] These AI-driven platforms offer innovative solutions that extend the reach of mental health services and provide continuous support and monitoring, distinguishing them from traditional therapy, which is limited by session availability.[15]

### 3.2 AI-Powered Tools for Diagnosis and Treatment

AI technologies are proving beneficial in the understaffed mental health care field by aiding in early diagnosis and creating personalized treatment plans. For instance, Dr. Christopher Romig, director of innovation at Stella, highlights the potential of AI to address the significant shortage of mental health providers in the U.S. [16] Companies like Click Therapeutics leverage AI algorithms to analyse patient data, enabling tailored treatment strategies for conditions such as depression and obesity.[16]AI applications also utilize sophisticated pattern recognition to identify early warning signs of mental health issues. By analysing vast datasets, including speech patterns and social media behavior, these systems can detect subtle changes indicating conditions like depression or PTSD, facilitating timely interventions.[15] Additionally, platforms like Winterlight Labs employ speech analysis to detect early signs of cognitive impairments, demonstrating the potential of AI to enhance diagnostic accuracy.[17]

### 3.3 Continuous Support and Accessibility

One of the significant advantages of AI applications is their ability to provide immediate assistance and feedback, overcoming barriers to therapy access. The anonymity and accessibility of AI-driven solutions create a stigma-free environment for users, encouraging them to seek help when needed.[15] This is particularly important for individuals who may hesitate to engage with traditional mental health services due to privacy concerns or logistical challenges.[15] Moreover, AI systems can monitor patients continuously, allowing for real-time adjustments to treatment plans based on user data. This capability not only supports dynamic care but also enhances the overall quality of mental health services provided.[15]

### 3.4 Ethical Considerations and Challenges

As AI technologies become increasingly prevalent in mental health care, they also raise ethical considerations regarding data security, confidentiality, and equitable access to services. Ensuring patient rights and maintaining transparency in AI applications are critical to their responsible integration into mental health practices.[3] While AI offers promising tools for improving mental health care, ongoing research and clinical validation are essential to establish their long-term efficacy and safety in diverse patient populations.[3]

## 4. Benefits of AI in Mental Health

The integration of artificial intelligence (AI) technology in mental health care is reshaping how individuals' access and experience support and treatment. AI offers a range of benefits that enhance both the reach and effectiveness of mental health services.

### 4.1 Increased Accessibility

One of the most significant advantages of AI in mental health is its ability to increase accessibility to care. AI-powered tools can provide support to individuals who may not have easy access to traditional mental health services, especially in underserved or remote areas. This democratization of care fosters equality and inclusivity in mental health support systems [9].

### 4.2 Early Detection and Diagnosis

AI algorithms are capable of analysing patterns in large datasets, including social media activity, medical records, speech patterns, and physiological data. This analysis can help identify potential signs of mental health issues at an early stage, offering timely intervention opportunities that may not be possible through traditional diagnostic methods [18][19].

### 4.3 Customized Recommendations

AI systems can tailor recommendations based on the specific symptoms or challenges faced by users. This personalization can include suggestions for therapy exercises, mindfulness techniques, or educational resources, thus enhancing the relevance of the care provided [20].

### 4.4 Adaptive Interventions

As users progress in their mental health journey, AI tools can adapt their recommendations to align with evolving needs. This ensures that the therapeutic approaches remain effective and relevant, enhancing the overall treatment experience [20].

### 4.5 Improved Patient Engagement

AI technologies can improve patient engagement by automating menial or repetitive tasks, thereby allowing clinicians to focus on more complex interactions. By streamlining administrative processes, AI can save costs associated with medical record keeping, prescription management, and insurance billing [18].

### 4.6 Enhanced Mental Health Tracking

AI is also revolutionizing mental health tracking through apps that integrate with wearable devices. These tools can monitor physiological indicators, such as heart rate and blood pressure, to assess changes in users' mental well-being. For instance, the BioBase app uses AI to interpret data from wearables and has reportedly reduced employee burnout, leading to fewer sick days [19].

### 4.7 Ethical Considerations and User Privacy

While the benefits of AI in mental health are substantial, it is crucial to navigate ethical considerations and prioritize user privacy. Concerns regarding patient privacy and informed consent must be addressed, especially in the context of AI applications that utilize sensitive data [10][19]. By ensuring ethical practices, the mental health sector can harness AI's potential to create a more effective and accessible care system for everyone [19].

## 5. Challenges and Limitations

### 5.1 Mitigation Strategies and Ethical Considerations

The integration of artificial intelligence (AI) in mental health care presents various challenges and limitations that must be carefully considered. One significant challenge involves the ethical implications of using AI in sensitive environments, such as psychiatric wards. While there are parallels between psychiatric inpatients and inmates, the complexities of mental health care require distinct approaches. In particular, the risks associated with involuntarily detained patients, those at risk of self-harm, and the potential for violence necessitate careful mitigation strategies [10]. As AI technology advances, it becomes crucial to explore its role in enhancing patient safety while ensuring ethical standards are maintained.

## 5.2 Limitations of Generative Language Models

Despite the promising applications of AI in mental health, several limitations remain. The technology's rapid evolution has not been matched by comprehensive guidelines or ethical frameworks, leading to concerns about accuracy, bias, and transparency [8][21]. For instance, generative language models, while increasingly accurate, can still produce confabulated outputs or "hallucinations," which may mislead both practitioners and patients. Furthermore, the quality and transparency of training data are critical issues, as biases inherent in these datasets can adversely affect AI outcomes and lead to inequities in care delivery [8][9].

## 5.3 Privacy and Consent Challenges

Another significant hurdle is the challenge of ensuring patient privacy and obtaining meaningful consent for data use. The sensitive nature of mental health data heightens privacy concerns, as unauthorized access or misuse of such information can have profound implications for individuals [4]. Although mechanisms like dynamic consent have been proposed to address these issues, the reliance on individual authorization can introduce biases, particularly if consent is not equally accessible to all patients [4]. Consequently, the ethical handling of health data remains a pivotal challenge as AI systems are developed and deployed in mental health contexts.

## 5.4 Developer Responsibilities and Ethical Frameworks

The responsibilities of developers in creating AI applications for mental health are also a point of concern. Given the potential vulnerability of individuals seeking mental health support, developers must adopt a care-centric approach throughout the design and implementation process. This approach emphasizes the importance of empathy, relational care, and ethical oversight, suggesting that ethical committees should be established to evaluate the impacts of AI tools in therapeutic settings [9]. By incorporating diverse perspectives and prioritizing the well-being of users, developers can better align AI innovations with the ethical standards of care.

## 6. Evidence and Research

### 6.1 Overview of AI in Mental Health

The integration of Artificial Intelligence (AI) into mental health applications has garnered significant attention in recent years, facilitating advancements in data collection, diagnosis, and the generation of automated interventions. This interdisciplinary field involves collaboration between technical experts and mental health professionals, addressing both the technological and therapeutic challenges within mental healthcare [21][3]. Despite the progress, the literature indicates that the collaboration between these domains often suffers from a lack of communication, with researchers in one field frequently unaware of developments in the other, leading to potential gaps in understanding and application [21].

### 6.2 Data Sources and Analysis

The utilization of Natural Language Processing (NLP) in mental health research has relied on a variety of data sources, including social media posts, blogs, and clinical documentation. These sources present distinct characteristics, such as variations in text length, vocabulary, and author intent, which are critical for effective analysis [5][21]. The collection and processing of this data have advanced considerably; however, there remains a notable deficit in research focusing on the applications of Natural Language Generation (NLG) in mental health interventions, particularly outside of specific cases like relational agents [4].

### 6.3 Ethical Considerations

Conducting research in the field of AI and mental health necessitates adherence to ethical guidelines and standards. Ethical approval for studies is typically required, as seen with the protocols established by the University of Melbourne's Human Research Ethics Committee [5]. Regulations such as the Common Rule ensure that research involving human subjects is rigorously reviewed to safeguard participant rights, while the Federal Trade Commission Act imposes restrictions against misleading health claims related to AI products [4]. These ethical frameworks are crucial to maintaining public trust and ensuring the responsible development of AI technologies in mental health settings.

### 6.4 Professional Training and Bias Mitigation

A critical aspect of ensuring accurate AI applications in mental health is the training of professionals in recognizing their biases. Previous studies have highlighted the lack of attention to the biases that annotators bring to the labelling and interpretation of data sets [13]. To create more representative and reliable datasets, it is essential for professionals to engage in reflexivity and be trained on their implicit biases, ensuring that diverse demographic groups contribute to data production [13][1].

## 6.5 Future Directions

The integration of artificial intelligence (AI) in mental health care is poised for substantial advancements that could reshape how individuals access support and treatment. As technology continues to evolve, several key areas warrant exploration for future development.

## 6.6 Personalized and Scalable Interventions

One of the most promising future directions is the development of personalized and scalable interventions using AI. By harnessing vast datasets and machine learning algorithms, mental health professionals can create tailored treatment plans that consider individual patient characteristics such as genetics, lifestyle, and prior treatment responses [12]. This personalization not only enhances the effectiveness of interventions but also fosters greater engagement from patients, ultimately leading to improved outcomes.

## 6.7 Enhancing Accessibility Through Technology

AI-driven solutions have the potential to significantly enhance accessibility to mental health care, especially for underserved populations. With innovations in app development and telehealth, individuals in remote or underserved areas can gain access to therapy and support that was previously unavailable [12]. This democratization of mental health resources is crucial as the global mental health crisis continues to escalate, affecting millions who lack traditional care options [17].

## 6.8 Integration with Emerging Technologies

The future of mental health AI also lies in its integration with emerging technologies such as Virtual Reality (VR) and wearable devices. VR can create immersive therapeutic experiences, particularly for conditions like PTSD and phobias, while AI can provide real-time feedback and adjust treatment parameters based on user interactions [20]. The fusion of these technologies can offer a more holistic approach to mental health care, emphasizing proactive self-care and immersive therapeutic methods.

## 6.9 Predictive Analytics for Early Intervention

Another critical area for future development is the use of predictive analytics to identify individuals at risk for mental health issues. AI algorithms can analyse patterns in data to flag potential crises, enabling timely interventions that could prevent the escalation of mental health conditions [22]. This proactive approach marks a shift from reactive to preventative mental health care, aligning with the broader goals of healthcare reform.

**6.10 Ethical Considerations and Responsible Development**

As AI technology continues to evolve, ethical considerations must be at the forefront of its application in mental health. Issues such as data privacy, algorithmic bias, and informed consent are paramount, requiring collaboration among AI experts, clinicians, and policymakers to ensure that AI applications are safe, equitable, and beneficial to all users [11]. Addressing these concerns will be essential to fostering public trust and ensuring the responsible deployment of AI in mental health services.

## 7. Conclusions

The integration of AI into mental health care represents a significant step toward revolutionizing the way psychological support and treatment are delivered. From its early origins with chatbot models like ELIZA to today's sophisticated machine learning systems, AI has the potential to enhance accessibility, personalize treatment, and enable real-time monitoring of mental health. This transformative development is particularly critical as mental health issues continue to escalate globally, highlighting the urgent need for innovative solutions that can address challenges such as access to care, the shortage of mental health professionals, and the stigma surrounding mental health treatment.

AI-powered applications, such as AI-driven chatbots, diagnostic tools, and continuous monitoring systems, offer immense promise in delivering efficient, scalable, and personalized mental health support. Notably, platforms like Wysa, Woebot, and Replika demonstrate the growing role of AI in providing scalable mental health care that transcends traditional in-person consultations. These AI systems not only provide immediate assistance but also allow for real-time adjustments to treatment plans, thus enhancing the overall quality and effectiveness of care. Moreover, AI's ability to analyse vast datasets—from speech patterns to social media behavior—enables early detection of mental health issues, offering timely intervention and personalized care tailored to individual needs.

However, despite its potential, the application of AI in mental health care raises significant ethical concerns. Privacy, data security, and the transparency of AI algorithms are critical issues that must be addressed to ensure the responsible integration of AI technologies into mental health practices. Algorithmic bias and the potential for inequitable access to care present further challenges that demand ongoing research, policy development, and regulatory frameworks to protect patient rights and promote fairness.

As the field evolves, it is imperative that AI technologies are deployed with robust ethical standards that prioritize the well-being of patients and foster equitable access to mental health care. AI must be seen not as a replacement for human providers, but as a complementary tool that enhances the capabilities of mental health professionals, expands access to care, and improves patient outcomes. With continued research, ethical vigilance, and collaboration between technologists and mental health professionals, AI has the potential to transform the landscape of mental health care, ensuring that support is accessible, personalized, and effective for all individuals, regardless of their geographical or socio-economic circumstances.

In conclusion, while AI offers substantial benefits, its successful integration into mental health care hinges on navigating the complex ethical, societal, and technological challenges it presents. By striking a balance between innovation and responsibility, we can harness the full potential of AI to create a more inclusive, effective, and sustainable mental health care system that meets the diverse needs of individuals worldwide.